# A Critical Review of Methods for Real-World Applications to Generalize or Transport Clinical Trial Findings to Target Populations of Interest

Albee Y. Ling[1], Maria E. Montez-Rath[2], Paulo Carita[3], Karen Chandross[4], Laurence Lucats[3], Zhaoling Meng[5], Bernard Sebastien[3], Kris Kapphahn[1], Manisha Desai[1]

[1]Quantitative Sciences Unit, Division of Biomedical Informatics Research, Department of Medicine, Stanford University School of Medicine, 1701 Page Mill Road, Palo Alto, CA 94304
[2]Division of Nephrology, Department of Medicine, Stanford University School of Medicine, 1070 Arastradero Road, Palo Alto, CA 94304
[3]Sanofi R&D Region, France
[4]Sanofi R&D Bridgewater, NJ, USA
[5]Sanofi R&D Cambridge, MA, USA
*Sanofi co-authors are listed in alphabetical order

# Abstract

Generalizability and transportability methods have been proposed to address the external validity bias of randomized clinical trials that results from differences in the distribution of treatment effect modifiers between trial and target populations. However, such studies present many challenges. We review and summarize state-of-the-art methodological considerations. We additionally provide investigators with a step-by-step guide to address these challenges, illustrated through a published case study. When conducted with rigor, such studies may play an integral role in regulatory decisions by providing key real-world evidence.

# 1. Introduction

Rigorous methods are needed to assess effects of experimental treatments, therapies, or interventions. Randomized clinical trials (RCTs) are the gold standard for evaluating treatment effects and are therefore leveraged whenever possible to describe the effect of the treatment in the real world. Well conducted RCTs can yield estimates with high internal validity, meaning that the measured effect reflects the true effect for the population studied (trial population). However, the study might lack external validity, meaning that the trial population might not be representative of other relevant populations of interest. More specifically, there may be a population of particular interest – referred to as the target population – that differs from that of the trial population. Such situations are common and can occur for a variety of reasons. For example, consider a trial established to assess efficacy of a new experimental regimen for breast cancer. Those designing the trial may want to assess this regimen among those that can really benefit. They define this patient population as those with Stage 2 or 3 primary breast cancer with tumors larger than 2cm that are estrogen-receptor (ER) and progesterone-receptor (PR) positive. Translating the findings from such a trial to the clinic then becomes a challenge because the trial population may not represent the typical breast cancer patient who comes into the clinic. If the latter is the target population of interest, more work needs to be done to understand the impact the regimen would have and whether clinicians should make recommendations about the regimen to their patients. Some may describe this phenomenon (lack of ability to translate findings of a trial to a target population) as a gap between treatment efficacy and treatment effectiveness, where the latter refers to the effect of the treatment in the "real-world".

More formally, when there are treatment effect modifiers whose distributions differ between trial and target populations, the treatment effect estimated for the trial population can be a biased estimate of the treatment effect in the target population [1–3]. For example, in the breast cancer example above, suppose the treatment effect is enhanced for those diagnosed at earlier stages and with smaller tumors and suppose the distribution of tumor size and stage in the trial sample differ from the target population of interest. In this case, the validity of the treatment effect estimated from the trial is threatened. When the trial population is a subset of the target population, we refer to the situation as a *generalizability* scenario. Consider again the breast cancer example. If the target population is breast cancer patients with Stage 2 or 3 disease, the trial population – those with Stage 2 or 3 disease, with a large enough tumor size and with ER and PR positivity status – is a subset of the target. In contrast, in a *transportability* scenario, the trial and target populations are disjoint [4–8]. For example, suppose we had studied the breast cancer regimen only among those diagnosed with Stage 3 disease (trial sample), as they were deemed most likely to benefit, and then wanted to know if the trial findings also applied to those diagnosed with Stage 2 disease (target population). There may be additional nuanced scenarios where there is some partial overlap between the trial sample and target population such as the case when the interest lies in transporting findings from a trial conducted on newly diagnosed Stage 2 patients to a target population of those with onset of metastases. While theoretically disjoint populations, in practice the data sources may have some overlap if, some diagnosed at Stage 2 also appear within the data resource of patients who had metastasized. Both study design and analytical solutions, such as *generalizability and transportability* studies, can be used to address this type of bias[5,9,10]. In one real example, investigators wanted to measure the effect of a highly active antiretroviral therapy in the target population of US people infected with HIV in 2006. A clinical trial had been conducted in 1,156 HIV-infected adults that concluded with positive results[1]. Thus, the authors proposed and described a generalizability study that repurposed the original clinical trial data and used data collected by the Centers for Disease Control and Prevention to define the target population of interest. Due to the high cost of conducting trials and the increasing number of observational data sources available, *generalizability and transportability* studies can be a valuable tool to generate or validate clinical hypotheses, provide real-world evidence for decision making, as well as facilitate future studies.

While there has been extensive methodological research for *generalizability and transportability* methods [1,2,15–24,3,25–29,4–6,11–14], the application of such tools is not always straightforward. For example, if using a propensity score (PS)-based method, how the model is specified for estimating the PS may have an impact on the findings. More specifically, for these methods to provide valid conclusions regarding the treatment effect in target populations, they rely on a series of assessments, including but not limited to key causal inference assumptions. Particularly problematic is a lack of understanding of when these methods can be appropriately applied given the assumptions needed for the validity of findings. Often, target populations are represented by real-world data sets which are known for being messy, noisy, or incomplete as their creation was often not intended for research. Thus, issues that arise in practice, such as missing data, unobserved covariates, and differences in measurement ascertainment can pose a further threat to violation of key assumptions. These issues and how they are handled can further compromise the rigor of these studies. Importantly, to go from evidence gathered from an RCT to evidence about the real-world requires receptivity of these studies by regulatory entities. For example, Health Technology Assessment (HTA) bodies review real-world evidence to inform critical decisions. Best practices and concrete guidelines for applying these methods in studies that translate findings to optimize receptivity by such regulatory entities are needed.

In this report, we set out to summarize the state-of-the-art considerations in translating clinical trial findings, i.e. applying *generalizability* or *transportability* methods (with an emphasis on PS reweighting methods), illustrated using examples from the applied literature (use cases) [30,31]. Key steps we touch upon are listed in **Table 1** and include 1) Appropriateness of a Study to Translate Findings; 2) Data Availability; 3) Identifiability Assumptions; 4) Generalizability and Transportability Methods; 5) Assessing Population Similarity; 6) Missing Data; 7) Sensitivity Analysis; and 8) Interpretation of Findings. Although our step-by-step guide follows the sequence above, these steps are not independent from one another, and investigators may need to iterate through several steps before finalizing the analysis plan. For example, if the trial and target population turn out to be sufficiently different, as described in Section 2.5, one may want to re-define the target population (discussed in Section 2.1). We conclude with steps investigators might want to take to plan such studies even at the trial design phase. It is our intention that this guide will not only help investigators conduct a *generalizability* or *transportability* study with much needed clarity and structure, but also show the potential value added by these real-world studies. In doing so, we hope to facilitate the engagement with regulatory bodies so that they are receptive to results from such studies.

**Table 1.** Workflow of Conducting a Generalizability and Transportability Study

| Study Steps | Important Considerations |
|---|---|
| **1. Appropriateness of a Study to Translate Findings** | Are there questions or concerns about a trial's external validity? Which target population is of interest to study? Is there any evidence that the trial results may not be applicable to the target population? Is this a *generalizability* or *transportability* study? Are endpoints translatable and meaningful to the target population? Which variables are potential treatment effect modifiers to include from a clinical perspective? |
| **2. Data Availability** | Are individual level data for the trial (patient characteristics, treatment, and outcome) available? Are individual level data for the target population (patient characteristics) available? |
|  | Are key patient characteristics measured in both the trial and target population data sources? Are they ascertained/reported similarly? If not, is there a way they can be harmonized?<br>Are the potential treatment effect modifiers in Step 1 measured and harmonizable in the trial and target populations? |
| **3. Identifiability Assumptions** | What statistical assumptions need to be met to carry out the study?<br>Are these assumptions feasible given your study design? |
| **4. Generalizability and Transportability Methods** | What statistical methods have been developed?<br>What are the pros and cons of each method?<br>Are there any statistical packages that facilitate their use? |
| **5. Assessing Population Similarity** | What metrics can be used to quantify the similarity between trial and target populations?<br>What criteria should be applied to determine if the two populations are sufficiently similar? |
| **6. Missing Data** | How much missing data are present in the variables proposed for use in the analysis?<br>Which missing data methods can be applied? |
| **7. Sensitivity Analysis** | Which sensitivity analyses should be conducted, particularly in case of unmeasured treatment effect modifiers? |
| **8. Interpretation of Findings** | How should findings from the *generalizability* or *transportability* study be interpreted and compared to the trial findings? How do the sensitivity analyses contribute to the principal interpretation? |

# 2. The Workflow

## 2.1 Appropriateness of a Study to Translate Findings

Due to stringent inclusion and exclusion criteria among other aspects of the trial study design, the trial and target populations may differ on the distribution of treatment effect modifiers. It is this particular difference that weakens the applicability of the trial findings to a target population of interest and that motivates pursuing a study to translate findings from the clinical trial to a target population. In these situations, we turn to the class of *generalizability and transportability* methods that can be applied after a trial is concluded to assess the finding's external validity. It is important to note that *generalizability and transportability* methods only address one source of bias that threatens external validity – bias due to the difference in distribution of effect modifiers between trial and target populations when there is treatment effect heterogeneity[29]. Such bias (sometimes referred to as "selection bias") is often due to strict inclusion and exclusion criteria in the trial and the willingness of eligible participants to take part in the study [30]. Other sources of external validity bias are listed in Figure 1 and include differences due to study setting, treatment, or outcome, and are assumed to be absent when considering a *generalizability* or *transportability* study [29]. For example, consider a study examining the effect of an antiviral among those newly diagnosed with COVID-19 who are outpatients, where the treatment is administered immediately (within 3 days of a positive PCR test). The effect of the drug on symptom resolution in the outpatient setting may not translate to the hospitalized patient. Importantly, a study that aims to translate these findings may not be appropriate here because of the presence of bias due to the heterogeneity in both the setting and in how the drug is administered. In the outpatient setting it is administered upon diagnosis (early in the disease) and in the hospitalized setting it is administered at hospitalization (which may or may not coincide with diagnosis, and more likely is after disease progression). Another example where such methods may not be appropriate due to the change in setting can be drawn from trials to assess the efficacy of at-home COVID-19 tests that are usually performed in a controlled environment where participants are supervised. Once the tests are released to the public there will be no supervision, which increases the likelihood of false negatives.

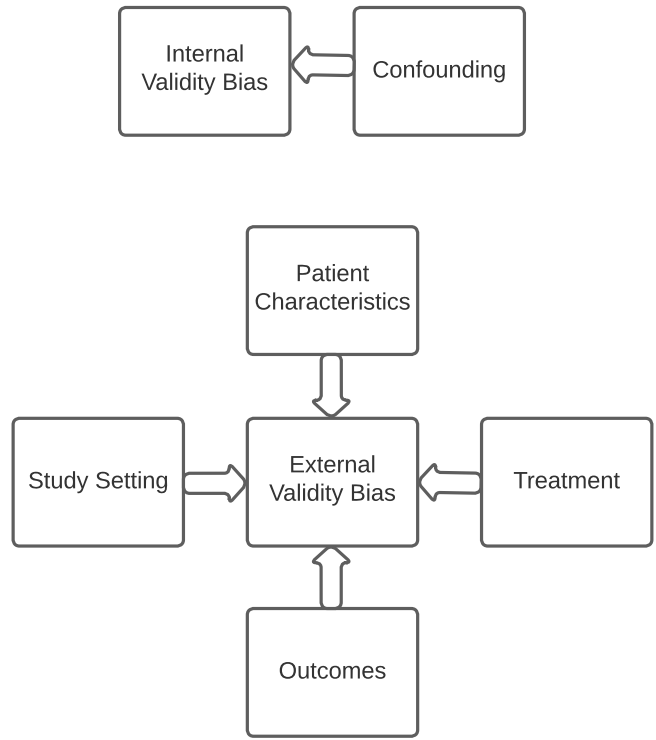

**Figure 1**. Sources of bias – measured or unmeasured – that threaten internal or external validity for a clinical study

As an example of an appropriate application of a study that translates findings and that we will continually refer to throughout this report, Susukida and others have previously compared 10 distinct trial populations in the National Institute of Drug Abuse Clinical Trials Network (CTN) to 10 respective target populations of patients drawn from the Treatment Episodes Data Set—Admissions (TEDS-A). They concluded that substance use disorder (SUD) patients recruited in CTN were not representative of a typical patient cared for at nationwide treatment facilities [30]. This motivated them to apply *transportability* methods to estimate the population average treatment effects in TEDS-A from CTN trial results. Specifically, to address a set of questions about interventions (10 questions) the trial population included 10 CTN trials conducted between 2001 and 2009 on three different broad types of interventions, each corresponding to a target population. We will use this example to illustrate our points throughout our report.

## 2.2. Data Availability
All the scientific considerations outlined in Section 2.1, Appropriateness of a Study to Translate Findings, must account for what data are available to the investigator. *Generalizability and transportability* methods generally require individual level covariate data on both trial and target populations, and treatment indicator and outcome variables in the trial population. If outcome variables existed in the target population, they can be used to check model specification or conduct an independent observational study [14]. Other methods have been developed to make use of summary level data when individual level data are not available [29]. Challenges can arise when individual trial data are not easily accessible, where there are limited variables in common between the trial and target data sources, and when variables are not measured comparably [11]. The choice of nine variables in Susukida et al. was made considering the overlap of patient characteristics in trial and target populations [31].

In addition to clinical knowledge, data-driven methods can also help identify variables that interact with treatment as potential effect modifiers [15,29]. To avoid leaving out important effect modifiers, it is always good practice to include as many potential treatment effect modifiers as possible based on clinical understanding and or prior evidence such as that obtained from subgroup analyses [11]. More recent studies indicate that investigators should prioritize variables that are part of the outcome generating functions [18]. For variables that are measured or ascertained differentially, investigators can re-code the variables, so they are comparable in trial and target populations.

The decision of which target population to use is usually driven by the research question of interest – it is typically more inclusive of the various types of subjects who are eligible to receive the interventions studied in the trial. All target populations were derived from the TEDS-A database using the respective eligibility criteria (e.g. age, time of study) of each trial [31]. Considering clinical relevance, meaning the variables relationship to outcome as well as potential differences in the populations, they included nine variables as potential treatment effect modifiers: sex, race/ethnicity, age, educational attainment, employment status, marital status, admission through criminal justice, intravenous drug use and the number of prior treatments for SUD.

## 2.3. Identifiability Assumptions

To illustrate principles, we formally describe the identifiability assumptions of a *generalizability* scenario in this section that can be easily extended to a *transportability* scenario. Let $X$ denote subject characteristics that are measured in both trial and target populations of interest, $T$ the treatment variable measured in the trial population only, and $Y$ the outcome variable required for measurement in the trial population. An indicator variable $S$ is assigned to each study subject, taking on the value of 1 if the subject is part of the trial and 0 otherwise. Under the potential outcomes framework [32], $Y_1$ and $Y_0$ are the two potential outcomes of an individual under treatment and control, respectively. Our research goal is to estimate the average treatment effect in the target population (PATE) as $E[Y_1 - Y_0]$. We have obtained an average treatment effect in the trial population (TATE) via $E[Y_1 - Y_0 \mid S = 1]$. When the trial sample is not a representative sample of the target population, TATE is a biased estimate of PATE.

The following assumptions need to be met before carrying out a *generalizability* study [11,14,25,29]. Due to the similarity between inverse probability of propensity score weighting (IPTW) and inverse probability of sampling (or selection) weighting (IPSW), these assumptions can be mirrored to conduct an observational study using IPTW and are as follows: <u>*positivity, exchangeability*</u>, and <u>*Stable Unit Treatment Value Assumption (SUTVA)*</u>. *Positivity* states that the probability of trial membership given the list of covariates lies between 0 and 1, or $0 < \Pr(S = 1 \mid X) < 1$. To illustrate using the Susukida example mentioned above, this means conditional on the list of nine covariates, each individual in the TEDS-A dataset could have been part of the CTN trials [31]. *Exchangeability* means that potential outcomes are independent of selection conditional on all covariates $Y^t \perp S \mid X$ for $t = 1$ and $t = 0$. This implies that, for all individuals in CTN trials or TEDS-A datasets, the outcomes occur in the same way, as a function of treatment and other variables including potential treatment effect modifiers. *SUTVA* means that the same version of treatment is given to all individuals in the trial and the target and that potential outcomes of a subject are not affected by other subjects (no interference).

An additional assumption that is unique to the class of *generalizability and transportability* studies, is that the range of values for each treatment effect modifier considered in the target population must be within the range of those in the trial population [20,33]. For example, suppose body mass index (BMI) is an effect modifier and a trial was conducted among morbidly obese adults (BMI > 35), so that the range of BMI values is 35 or greater. In this case, it is not feasible to apply *generalizability* or *transportability* methods to a target population of those at-risk of obesity where the range of BMI values would not include values over 35.

Several other implicit assumptions must also hold, including 1) *consistency*: if $T_i = t$, then $Y_i^t = Y^{T_i} = Y_i$ for individual $i$, meaning that the individual's potential outcome is the one actually observed under treatment *t*; 2) internal validity of the trial (the treatment effect is unbiased for the trial population); and 3) no missing data or measurement error on covariate, treatment, or outcome variables and 4) all scientist models are specified correctly. Under all the identifiability conditions mentioned above, TATE and PATE can be estimated by $E[Y_1 - Y_0 \mid S = 1] = E[Y_1 \mid S = 1] - E[Y_0 \mid S = 1]$ and $E[Y_1 - Y_0] = E[Y_1] - E[Y_0]$, respectively. Further, *generalizability* methods can be applied to estimate PATE.

## 2.4. Generalizability and Transportability Methods

There are three broad classes of *generalizability and transportability* methods: inverse probability of sampling (or selection) weighting (IPSW) [1,13,14,16,27,34], outcome model-based approaches [14,22,33,34], and a hybrid of both [14,22,33,35]. IPSW is an adaptation of inverse probability of propensity score weighting (IPTW), in which PS is defined as the probability of being in the trial population, conditional on pre-treatment covariates. PS is estimated in the entire target population (for *generalizability* scenarios) or the super-population where trial and target populations are concatenated (for *transportability* scenarios). Various modeling techniques can be used to model PS, including logistic regression or random forest. Weights are estimated based for the trial population only, as the inverse of PS for *generalizability* and inverse of odds for *transportability* [6,27].

$$\text{Generalizability: } w_i = \begin{cases} \frac{1}{P(S_i = 1 \mid X_i)}, & S_i = 1 \\ 0, & S_i = 0 \end{cases}$$

$$\text{Transportability: } w_i = \begin{cases} \frac{P(S_i = 0 \mid X_i)}{P(S_i = 1 \mid X_i)}, & S_i = 1 \\ 0, & S_i = 0 \end{cases}$$

Next, outcomes in the trial population are weighted so the distribution of baseline covariates resemble that of the target population [1,13,14,16,27,34]. Depending on the type of outcome, regression techniques (such as linear regression, logistic regression, or Cox proportional hazards models) can be modified to use the weights above to obtain PATE. Similarly to IPTW, techniques such as weight trimming, stabilized weights [36], or standardization [14] can be applied to handle large or unstable weights. Researchers have used both bootstrap and robust sandwich estimators to estimate variance [29].

Susukida et. al applied IPSW, to their study, where PS was estimated using random forest based on the nine variables common to trial and target populations. Linear and logistic regression with *transportability* weights estimated based on PS were performed to estimate PATE for each of the CTN trials, with 95% confidence intervals presented using robust standard error estimates.

In outcome model-based approaches, each potential outcome is modeled separately using trial data. The models are then used to predict the potential outcomes in the target population, from which PATE can be obtained by averaging the individual causal effects obtained from the difference of predicted potential outcomes [14,22,33,34]. In addition to regression methods, other machine learning methods such as Bayesian Additive Regression Trees (BART) are also commonly applied [34]. Compared to IPSW, they perform particularly well when the covariates are strong predictors of the outcome. The hybrid approaches combine IPSW and the outcome-based approaches, involve more flexible modeling, and are doubly robust under certain conditions [14,22,33,35].

In simulations where both selection model and outcome model are linear, all three classes of methods performed well in terms of bias when the models were correctly specified; outcome based methods had the lowest variance, followed by the hybrid approaches, the re-weighting methods, which had the largest variance [14]. Rudolph and others have shown that doubly robust estimators outperformed IPSW in terms of mean squared errors even under model misspecification [37]. A more comprehensive simulation study found that the relative performance of the various estimators also depends on the linearity of the outcome generating function and the target sample size and/or trial to target ratio [23]. For example, no estimators achieved satisfactory results when the trial was 2% of the target population, which was smaller than 30,000 individuals [23]. Relevant R packages include *generalize* [33], *dbarts* [38], *tmle* [39], and others depending on the specific models to be implemented [23].

## 2.5. Assessing Population Similarity

It is important to quantify the similarity between trial and target population before finalizing the analysis plan [24]. An adequate level of similarity provides greater confidence in generalized or transported findings. Standardized $\Delta p$ has been used by Susukida et al., and it is estimated by dividing the difference in mean PS between trial and target population by the pooled standard deviation of PS [2]. Lower values of $\Delta p$ indicate greater similarity between two populations, although different investigators may have their own preferences in choosing a specific threshold [30]. Among the 10 CTN trials, Standardized $\Delta p$ ranged from 1.06 to 2.08, suggesting a large difference between the two populations [30].

Another PS-based measurement, Tipton index, has also been used as a similarity metric and is defined as $\int \sqrt{f_s(s) f_p(s)} \, ds$, where $f_s(s)$ and $f_p(s)$ denote the distributions of PS in trial and target populations respectively [26]. Tipton index makes no assumptions about the PS distribution and takes on values between 0 and 1. The interpretation of Tipton index is detailed below (**Table 2**), based on simulation studies[26].

As with PS methods used in the context of observational studies, standardized mean difference (SMD) can also be a useful tool here[40]. Unlike standardized $\Delta p$ and the Tipton index, SMD measures the similarity of individual covariates. The comparison of SMD before and after weighting can also be used as a diagnosis tool for tuning the PS model and assessing how well the model adjusted for the difference in individual baseline characteristics in the two populations. While some common thresholds exist for assessing the covariate balance, there is no general consensus, and it is at the investigator's discretion to interpret the values [11].

**Table 2**. Interpretation of the Tipton Index

| Tipton index | Category | Interpretation |
| --- | --- | --- |
| [0.9, 1] | Very high | Generalizable, trial is very close to a random sample from target population |
| [0.8, 0.9) | High | Generalizable, very small bias and increase in standard error due to reweighting |
| [0.5, 0.8) | Medium | Generalization is possible, but with some bias and inflated standard errors |
| [0, 0.5) | Low | Not generalizable |

In case the difference between trial and target populations are too big, one can refine the research question, and consider re-defining the target population to be more similar to the trial while clarifying the relationship between this newly defined target population and the original target [25,29].

## 2.6. Missing Data

Missing data is an important aspect in the applications of *generalizability and transportability* methods. Missingness is common in both trial and target populations, and the level of missingness is often not trivial. For example, in the study by Susukida et al., the percentage of patients missing at least one variables could be as high as 10.4% among the 10 trials and 85.3% among the target populations [30]. Among the variables collected within a trial, the level of missingness could be as high as 71% (number of prior treatments in CTN01 trial)[30]. Only a handful of authors chose to apply multiple imputation (MI) in their study to translate findings[14,30,34,41–43], while most used complete cases analysis [44,45] or single imputation [11]. Although the key assumption of MI may not always hold (e.g., data were missing at random (MAR) conditional on observed variables), MI has been proven to have excellent statistical properties and further properties that are superior to that of complete case and single imputation. Importantly, the missing completely at random (MCAR) assumption – where missingness is unrelated to both observed and unobserved variables – is relied upon when performing a complete case analysis and is less flexible than the MAR assumption made under MI.[46] Susukida and others set a great example by being meticulous about reporting and characterizing missing data and describing missing data methods. They adopted an MI approach that we refer to as *MI-passive* (estimate PS after imputing underlying PS variables), and an approach for integrating imputation results referred to as *PSI-across* (averaging PS across all multiply imputed datasets), and included all available covariates from both trial and target populations in their imputation model [30,31]. Based on simulation results by us and others [47–50], we recommend implementing *MI-passive*, *PSI-within* (conduct IPSW within each imputed dataset), and additionally include trial indicator (for both trial and target populations), treatment and outcome variables (among trial participants) in the imputation model. Alternatively, we would also recommend coupling MI with bootstrapping methods to estimate the uncertainty of the treatment effect. Another great example is set by Mollan et al. who used a bootstrap variance estimator together with MI to handle missing data [43].

## 2.7. Sensitivity Analysis

When some of the assumptions mentioned in Section 2.3 are difficult to assess or justify, sensitivity analyses can provide more insight into the impact assumption violations have on the effect estimates. For example, Thabane et al [51] listed the following critical areas that should be considered in order to express the level of confidence in the conclusion:

- Will the results change if I change the definition of the outcome (e.g., using different cut-off points)?
- Will the results change if I change the method of analysis?
- Will the results change if we take missing data into account? Will the method of handling missing data lead to different conclusions?
- How much influence will minor protocol deviations have on the conclusions?
- How will ignoring the serial correlation of measurements within a patient impact the results?
- What if the data were assumed to have a non-Normal distribution or there were outliers?
- Will the results change if one looks at subgroups of patients?
- Will the results change if the full intervention is received (i.e., degree of compliance)?

Thus far in the literature of *generalizability and transportability* method applications, sensitivity analyses typically consisted of alternative analytical choices that were specific to the data example. Some examples are imputing missing data based on different assumptions [44], varying the variables included as potential effect modifiers [45], and accounting for RCT dropouts [43]. Depending on the research question and data availability, we highly recommend investigators conduct such analyses and following the considerations of Thabane and others [51]. They are not limited to the ones mentioned above and can include potentially re-defining a target population that is more like the trial population (Section 2.5) and alternative *generalizability* or *transportability* methods (Section 2.4) (e.g., IPSW or outcomes-based methods).

Methodological research for performing sensitivity analyses has been developed to address situations where certain effect modifiers are only observed in the trial but not in the target population, and when certain effect modifiers are unobserved in either population[19,20], or more generally when the exchangeability assumption is violated [52]. For example, imputation-based techniques can be used to evaluate robustness of findings under various scenarios of unobserved effect modifiers. While we highly recommend the incorporation of sensitivity analyses when interpreting findings, these methods have not been widely adopted in the applied literature yet.

## 2.8. Interpretation of Findings

A natural component of the interpretation of studies that translate findings to target populations is to compare the effects between the trial and target populations. We found in the literature that treatment effects estimated using *generalizability and transportability* methods were compared to those estimated in the trial sample in terms of magnitude, direction, and statistical significance [1,13,31,33]. Incorporating these existing considerations in the literature, we propose comparisons that cover three key areas: [53]. 1) *Regulatory agreement:* The two treatment effects are in regulatory agreement if they agree both on the direction and significance [53], 2) *Estimation agreement:* they are in estimation agreement if the average treatment effect in the target population falls into that of trial population [53] and 3) *Design agreement:* they are in design agreement if the estimated treatment effect fulfills a trial pre-specified threshold. For example, the SPRINT trial was designed so that a significant effect would provide a 20% benefit. If the average treatment effect estimated in the target population was also larger than 20%, we can conclude that they are in design agreement [54]. Standardized differences can also be used to quantify the difference between two treatment effect estimates [53]. This can be accomplished by dividing the difference between the two treatment-effect estimates by their pooled standard error. Importantly, one needs to be cautious in interpreting the results considering limitations due to missing data, unmeasured effect modifiers, or any imbalance in measured effect modifiers after weighting. Thus, the choice of missing data methods and sensitivity analysis is critical when drawing conclusions.

To interpret the results of Susukida et. al using the three metrics proposed by Franklin et al. [53], we devised the following table below (**Table 3**). For example, both CTN7 and CTN10 reached both regulatory and estimate agreement, with very small, standardized difference (-0.31 and 0.29 respectively). We elaborate on CTN7 here – the trial was conducted from 2001 to 2003, where 388 participants, using cocaine or methamphetamine and entering a substance abuse treatment program, were randomized to receiving incentives (along with standard care therapy) or treatment as usual. The two primary outcomes were 1) percent of submitted urine cocaine, amphetamine and methamphetamine-free, and 2) longest duration of abstinence from primary target drugs (cocaine, amphetamine, methamphetamine, alcohol) [55]. The transported results mean that the treatment effect was statistically significant in both trial and target population, that the PATE falls into the 95% confidence interval of trial results, and that the two treatment estimates are comparable as quantified by standardized difference. It is up to the investigator to determine if this means that the trial results *transport,* and such interpretation might vary depending on the purpose of the study (e.g., for academic study that provides further insight or submission to FDA for approval of the treatment for a particular indication).

**Table 3.** Susukida et. al Results and Interpretation. TATE = average treatment effect in the trial; PATE = average treatment effect in the population; CI = confidence interval; SD = standard deviation.

| Trial | TATE (95% CI) | TATE SD | PATE (95% CI) | PATE SD | Regulatory Agreement | Estimate Agreement | Standardized Difference |
|---|---|---|---|---|---|---|---|
| CTN1 | 6.47 (1.60, 11.35) | 2.48 | 0.58 (-3.82, 4.98) | 2.24 | No | No | -1.76 |
| CTN2 | 3.07 (-1.77, 7.90) | 2.47 | 13.10 (5.82, 20.37) | 3.71 | No | No | 2.25 |
| CTN3 | 0.63 (-1.75, 3.00) | 1.21 | 3.92 (-1.31, 9.15) | 2.67 | **Yes** | No | 1.12 |
| CTN4 | -2.52 (-4.26, -0.79) | 0.89 | -3.02 (-6.98, 0.94) | 2.02 | No | **Yes** | -0.23 |
| CTN5 | -0.84 (-2.88, 1.20) | 1.04 | 1.31 (-5.57, 8.20) | 3.52 | **Yes** | No | 0.59 |
| CTN6 | 0.16 (-1.36, 1.68) | 0.78 | 2.53 (-0.34, 5.41) | 1.47 | **Yes** | No | 1.43 |
| CTN7 | 0.26 (-1.34, 1.87) | 0.82 | -0.12 (-1.89, 1.66) | 0.91 | **Yes** | **Yes** | -0.31 |
| CTN10 | -0.94 (-5.44, 3.57) | 2.30 | -3.38 (-5.57, -1.19) | 1.12 | No | **Yes** | -0.96 |
| CTN13 | 0.72 (-2.35, 3.78) | 1.57 | 1.70 (-4.06, 7.46) | 2.94 | **Yes** | **Yes** | 0.29 |
| CTN30 | -1.79 (-3.37, -0.20) | 0.81 | 0.85 (-4.08, 5.78) | 2.52 | No | No | 1.00 |

Post hoc subgroup analyses using trial data can also be carried out to aid the interpretation of results, acknowledging its limited power in detecting effect heterogeneity [11,31]. For example, in cases where there was no agreement between TATE and PATE in terms of statistical significance, one can provide insight into where heterogeneity of treatment effects may play a role by describing the effects in the respective sub-groups. Susukida et. al made thoughtful choices when conducting such analyses: 1) they only focused on scenarios where the TATE and PATE did not agree in terms of statistical significance, in other words, when they do not reach regulatory agreement; and 2) they only investigated the covariates whose distribution differed significantly between trial and target.

### 2.9. Future Study Design Considerations

Ideally when designing the trial, the trial sample will have the same distribution of effect modifiers as that in the target population. This can be accomplished by designing the trial to be a random sample of the target population or by performing stratified sampling based on prognostic factors [5]. Although designing a trial as such is preferred[5,29], *generalizability and transportability* studies may still be of interest if these trial design options are not feasible, if more than one target populations are of interest, and if target populations change over time [5]. There are a few considerations to facilitate future *generalizability and transportability* studies that should be weighed in the trial design phase. For one, the researcher should pre-specify the target population(s) and adopt appropriate sampling techniques when designing the trial[10]. Recruitment records can also be analyzed post-trial to understand the difference between trial and intended target populations [56]. For another, data on potential effect modifiers should be collected on trial participants preferably in the same format as possible data resources that could be used to define the target population.[5] Otherwise, special attention will be required in order to harmonize data to generalize or transport findings[8]. To that end, whenever possible, trialists could plan ahead to link trial data to observational databases that capture the target population[57]. Additionally, investigators should keep in mind that trials with limited sample sizes will produce large confidence intervals when *generalized* or *transported* to a larger target population (a common scenario). PATE estimates will naturally have larger standard errors than TATE estimates because of the larger amount of uncertainty from incorporating weighting techniques [11]. As such, trials should be designed to detect the desired effect estimate on the trial sample while keeping in mind the uncertainty of future studies conducted to demonstrate the external validity of the trial.

## 3. Conclusion

We have summarized important methodological considerations when conducting *generalizability and transportability* studies to translate clinical trial findings to target populations of interest. Moreover, we have provided concrete examples to consolidate the theory discussed including a published case study to illustrate the methods. As we have emphasized throughout, investigators are strongly encouraged to discuss the plausibility of key assumptions (coupled with results from sensitivity analyses that challenge such assumptions) and limitations of their specific study when interpreting findings and drawing conclusions.

Additionally, the promise of these studies is growing and has additional potential applications. For example, one can think of the *generalizability and transportability* methods described above as a means to standardize trial sample to resemble the target population[1]. In this manner, they can also be combined with other types of studies (e.g. a comparative effectiveness cohort study [58–60], estimating incidence of events [61–63]) as a tool of standardization.

With observational data becoming increasingly available, *generalizability and transportability* studies will be a great source of RWE. Such RWE can be used in conjunction with other evidence generated from clinical trials, pragmatic trials, and other observational studies, where internal and external validity are considered jointly [6,28]. Investigators will need to be diligent about reconciling any differences among these studies by considering study design and other practical constraints (e.g. data availability) to reach a conclusion that is clinically meaningful [7].

## Acknowledgements

We are graciously funded by an iDEA Award (https://www.sanofi.us/en/innovation-and-science/partnering-initiatives/sanofi-idea-awards)